\documentclass[11pt]{article}
\usepackage{style}
\usepackage{shortcuts}
\usepackage[normalem]{ulem}

\title{The Hyperclique Hypothesis is False}
\author{Nick Fischer\thanks{Max Planck Institute for Informatics, Saarbrücken, Germany. Email: \href{mailto:nfischer@mpi-inf.mpg.de}{nfischer@mpi-inf.mpg.de}.}}
\date{}

\begin{document}
\maketitle

\begin{abstract}
The Hyperclique Hypothesis postulates that, for all constants $k > h \geq 3$, detecting a $k$-clique in an $h$-uniform hypergraph with~$n$ vertices requires time $n^{k-o(1)}$. This hypothesis is false: there is an algebraic algorithm that detects (and even counts) $k$-cliques in $h$-uniform hypergraphs in time $O(n^{(1-\epsilon_h)k})$. Via known reductions, this also refutes the recently proposed $k$-AP Hypothesis. As a consequence, many problems with conditional lower bounds based on these two hypotheses may admit faster algorithms after all.  This result is almost entirely due to AI.
\end{abstract}

\vspace*{-.5ex}
\begin{ai}
The results of this paper were obtained in a lengthy chat with ChatGPT Astra~6 (OpenAI) and then refined by Claude Opus~5.5 (Anthropic), in both cases with only minimal human input. Both models also assisted in preparing this write-up. I take full responsibility for the contents of this paper.
\end{ai}

\thispagestyle{empty}
\setcounter{page}{0}
\clearpage

\section{Introduction}
The goal of fine-grained algorithm design and complexity theory is to pinpoint the best possible running time of every computational problem. On the one hand, this involves designing ever faster algorithms. On the other hand, it involves proving \emph{conditional lower bounds}, that is, showing that a problem cannot be solved faster than currently known unless some established hardness hypothesis fails. Over the past fifteen years, this approach has developed into a major success story, connecting dozens of polynomial-time problems to a small set of canonical hardness assumptions; see, e.g.,~\cite{Williams19}. The reasoning behind this approach is not that we treat hypotheses as dogmas, but rather that they mark the current frontier of our algorithmic knowledge: if a problem cannot be solved faster without refuting some hypothesis, then one should \emph{first} try to refute that hypothesis. In other words, the hypotheses show us where to concentrate our efforts. 

We now seem to have entered a new era in which the algorithmic frontier is moving dramatically. Powered by AI systems, several central hardness hypotheses have been refuted within a matter of weeks, in each case by an algebraic algorithm. Among them are the 3SUM Hypothesis~\cite{GajentaanO95} and the APSP Hypothesis~\cite{WilliamsW18}, both refuted by Alman and Vassilevska Williams~\cite{AlmanW26} (see also~\cite{KaringulaL26}), as well as the Exponential Time Hypothesis (ETH)~\cite{ImpagliazzoP01,ImpagliazzoPZ01}, refuted independently by OpenAI~\cite{OpenAI26b} and Anthropic~\cite{Claude26}. Again, this should not be viewed as discouraging, but as exciting progress: the algorithmic frontier has finally moved, and we seem ready to design faster algorithms for many problems.

This paper adds another hypothesis to this list: the \emph{Hyperclique Hypothesis}, proposed by Lincoln, Vassilevska Williams, and Williams~\cite{LincolnWW18}. It states that for all constants $k > h \geq 3$ and $\epsilon > 0$, there is no $O(n^{k-\epsilon})$-time algorithm for detecting a $k$-clique in an $h$-uniform hypergraph with $n$ vertices, that is, a set of $k$ vertices any $h$ of which form a hyperedge. The restriction to $h \geq 3$ is crucial. For $h = 2$, the problem degenerates to $k$-clique detection in graphs, which can be solved in time~$O(n^{\omega k/3})$ (when $k$ is divisible by $3$) using fast matrix multiplication~\cite{NesetrilP85}; here, $\omega \leq 9/4$ is the exponent of matrix multiplication~\cite{OpenAI26a}. For $h \geq 3$, in contrast, no algorithm polynomially faster than exhaustive search was known. The main result of this paper is an algebraic algorithm showing that the Hyperclique Hypothesis is false, for every choice of constants~$k > h \geq 3$.

\begin{theorem}[Hyperclique Detection] \label{thm:hyperclique}
For every constant $h \geq 2$, there is a constant $\epsilon_h > 0$ such that, for every constant $k > h$, a $k$-clique in an $h$-uniform hypergraph with $n$ vertices can be detected in time $O(n^{(1-\epsilon_h) k})$.
\end{theorem}

\subsection{Consequences}
Over the past years, many conditional lower bounds have been based on the Hyperclique Hypothesis. They concern problems as diverse as Orthogonal Vectors in moderate dimension~\cite{AbboudBDN18}, sparse cycle and hypercycle detection~\cite{LincolnWW18,FuLR25}, induced $4$-cycle detection~\cite{DalirrooyfardW22}, approximate $n$-pairs shortest paths~\cite{DalirrooyfardJWW22}, subgraph isomorphism for patterns of bounded treewidth~\cite{BringmannS21}, lowest common ancestors in DAGs~\cite{MathialaganWX22}, first-order model checking and Boolean constraint satisfaction problems~\cite{BringmannFK19,KunnemannM20,AnGIJKN21,BringmannCFK22,FischerKRS25,FritschKRS26,FischerKR27}, geometric problems like Klee's measure problem~\cite{Kunnemann22,GorbachevK23}, diameter computation in geometric intersection graphs~\cite{BringmannKKNP22,ChanCGKLZ26a,ChanCGKLZ26b}, discrete $k$-center~\cite{ChanHY23}, and $L_\infty$ Hausdorff distance under translations~\cite{AngrickBGK26}, arithmetic problems including 4-SUM~\cite{GokajK25}, and some others~\cite{WilliamsX20,Zamir23}. These lower bounds have now lost their basis, so for many of the affected problems we can again hope for faster algorithms than currently known.\footnote{Some lower bounds in database theory, such as those in~\cite{CarmeliZBKS20}, rely on a weaker hypothesis that only rules out near-linear time in the number of hyperedges; this paper does not refute that hypothesis.}

For several problems, \cref{thm:hyperclique} even yields faster algorithms via known reductions. Most notably, for every $h \geq 3$, Max-$h$-SAT on $n$ variables can be solved in time $O^*(2^{(1-\delta_h) n})$ for some constant $\delta_h > 0$, and the same holds for Boolean Max-CSPs whose constraints have degree at most~$h$~\cite{Williams07,LincolnWW18}; previously, no such improvement over exhaustive search was known for any $h \geq 3$. In fact, the complexity of Max-$h$-SAT has also been suggested as a hardness hypothesis in its own right, including for sparse instances~\cite{AlmanW20}. Moreover, Zero-Weight and Min-Weight $k$-Clique in $h$-uniform hypergraphs can be solved polynomially faster than exhaustive search for every \makebox{$k > 2h$}~\cite{AbboudBDN18}; for $h=2$ this already follows from~\cite{AlmanW26}. Further consequences include faster algorithms for certain weight-$k$ Boolean CSPs, such as Densest $k$-Subhypergraph and Partial Vertex Cover in hypergraphs~\cite{KunnemannM20,FischerKRS25}, and polynomially faster algorithms for $(k,h)$-Orthogonal Vectors~\cite{FischerKR27} for $k>h$ with applications to certain first-order optimization problems and Boolean CSPs~\cite{BringmannCFK22,FischerKRS25}. Finally, the algorithm behind \cref{thm:hyperclique} is general enough to count the induced or non-induced copies of any fixed $k$-vertex pattern in an $h$-uniform hypergraph in time $O(n^{(1-\epsilon_h) k})$ whenever $k>h$.

In addition, \cref{thm:hyperclique} refutes the \emph{$k$-AP Hypothesis}, which was recently proposed by Dudek, Fischer, Gokaj, Jin, K\"unnemann, Mao, and Red\v{z}i\'c~\cite{DudekFGJKMR26}. It states that for every constant~\makebox{$k \geq 4$}, there is no $O(N^{2-\epsilon})$-time algorithm (for any constant $\epsilon > 0$) for detecting a nontrivial $k$-term arithmetic progression ($k$-AP) in a given set $A \subseteq \set{0, \dots, N}$. The $k$-AP Hypothesis is known to imply the Hyperclique Hypothesis~\cite{DudekFGJKMR26}: if a $k$-clique in a $(k-1)$-uniform hypergraph can be detected in time $T(n)$, then a $k$-AP can be detected in time $O(T(N^{2/k}))$. Combining this reduction with \cref{thm:hyperclique} for $h = k-1$ yields the following algorithm.

\begin{theorem}[$k$-AP Detection] \label{thm:kap}
For every constant $k \geq 3$, there is a constant $\epsilon_k > 0$ such that a nontrivial $k$-AP in a given set $A \subseteq \set{0, \dots, N}$ can be detected in time $O(N^{2-\epsilon_k})$.
\end{theorem}

Consequently, the problems that were shown to be hard under the $k$-AP Hypothesis are also open again. These include verifying certain identities in finite algebraic structures~\cite{DudekFGJKMR26} and many of the formerly 3SUM-hard geometric problems~\cite{Fischer26}.

\subsection{Circumventing the Algebraic Barriers} \label{sec:barriers}
Before this paper, two barriers suggested that algebraic techniques perhaps cannot speed up hyperclique detection. This subsection explains both barriers in elementary terms, as well as how the new algorithm avoids them.

For starters, consider the matrix multiplication problem. Up to constant factors, computing a matrix product~$AB$ is as hard as evaluating the following trilinear polynomial~\cite{BaurS83}:
\begin{equation*}
    H_2(A, B, C) = \sum_{x, y, z \in [n]} A_{xy} B_{yz} C_{xz}.
\end{equation*}
For trilinear polynomials like this, the complexity of algebraic algorithms, which may only add, subtract, multiply, and possibly divide numbers~\cite{Strassen73}, is captured by the so-called \emph{tensor rank}~$r$. The reason is that any such algorithm can be rearranged into a rigid form which only computes~$r$ products of three linear forms, (linear combination of the entries of~$A$) times (linear combination of the entries of $B$) times (linear combination of the entries of $C$), and adds up these products. Under this characterization, designing fast algebraic algorithms for trilinear polynomials is equivalent to proving upper bounds on their tensor rank. This perspective underlies much of algebraic algorithm design, starting with Strassen's seminal matrix multiplication algorithm~\cite{Strassen69}; see, e.g., the survey~\cite{Blaser13}.

Our goal, instead, is to evaluate the $4$-linear \emph{hyperclique polynomial}
\begin{equation*}
    H_3(A, B, C, D) = \sum_{x, y, z, w \in [n]} A_{xyz} B_{xyw} C_{xzw} D_{yzw}.
\end{equation*}
Its tensor rank (and even its border rank) is known to be~$n^4$~\cite{LincolnWW18}, which matches the trivial decomposition into its $n^4$ terms. Thus, the rank-based approach alone, which multiplies four linear combinations of the entries of~$A$,~$B$,~$C$, and~$D$, respectively, cannot beat exhaustive search. This is the first barrier.

However, this does not rule out fast algorithms, since for $4$-linear polynomials the precise correspondence between algorithms and tensor rank breaks down; see, e.g.,~\cite{BrandCKLOSW26,BrandKW26}. Algorithms now have more freedom; in particular, they can multiply two intermediate results that each depend on two of the inputs. The simplest example is the polynomial $(\sum_i a_i b_i) \cdot (\sum_j c_j d_j)$, which can be evaluated with $2n + 1$ multiplications even though its tensor rank is~$n^2$. Hence, to beat exhaustive search, an algorithm must form intermediate products that involve just two of the inputs, say, products of entries of~$A$ and~$B$. A natural way to organize this is to fix an evaluation order, such as $(AB)(CD)$, which determines in which order the inputs are combined and which variables are grouped together. This strategy is formalized by the model of \emph{tensor networks} studied by Austrin, Kaski, and Kubjas~\cite{AustrinKK22}, where we fix the evaluation order through a tree and allow rank-based algorithms for combining inputs or intermediate results. This model is strictly more powerful than tensor rank alone. It captures, for example, the fastest known algorithms for counting $k$-cliques in graphs~\cite{NesetrilP85}, which group the $k$ vertices into three blocks, although the corresponding polynomial has maximal tensor rank (and even maximal border rank)~\cite{AustrinKK22}. Nevertheless, Austrin, Kaski, and Kubjas proved that every tensor-network algorithm for~$H_3$ takes time~$\Omega(n^4)$. This is the second barrier.

To circumvent this second barrier, an algorithm should evaluate different parts of the polynomial in different orders, say, some parts as $(AB)(CD)$ and others as $(BC)(DA)$. The new algorithm does exactly that. It first keeps rearranging~$H_3$ as a signed sum of many smaller polynomials. It then evaluates each of these polynomials with whichever of the two pairings $(AB)(CD)$ or $(BC)(DA)$ is cheaper. Intuitively, each pairing is cheap for some of the terms, but neither is cheap for all of them. In other words, every single term is evaluated within the tensor-network model; the gain comes from \emph{mixing} different evaluation orders for different terms.

Interestingly, this strategy is closely related to Anthropic's recent refutation of the Exponential Time Hypothesis~\cite{Claude26}, which also rewrites a count as a signed sum of terms, chooses an efficient evaluation order separately for each term, and recurses. In fact, the two approaches are also connected on a more technical level (see the discussion after \cref{lem:swap}). OpenAI's independent ETH refutation~\cite{OpenAI26b} also follows a conceptually similar idea.

\paragraph{Further Related Work}
The algebraic complexity of hyperclique has also been considered in two very recent works. In~\cite{BrandCKLOSW26}, Brand, Curticapean, Kaski, Li, Orzel, Seppelt, and Wang identify obstacles to composing arithmetic circuits, assuming the Hyperclique Hypothesis. In~\cite{BrandKW26}, Brand, Kaski, and Wang introduce the \emph{two-slice rank}, another algebraic measure that captures mixed evaluation orders as above and thus goes beyond tensor networks. They show that it lower-bounds the number of multiplications needed to evaluate a polynomial, and propose proving that the two-slice rank of~$H_3$ is~$\Omega(n^4)$ as a route towards algebraic hardness of hyperclique. Instead, this paper shows that it is~$O(n^{4-\epsilon})$.

\section{Preliminaries} \label{sec:prelims}
\paragraph{Notation}
We write $[n] = \set{1, \dots, n}$, and $\binom{X}{h}$ for the set of all $h$-element subsets of a set~$X$. For a tuple $x = (x_1, \dots, x_k)$ and a set $S = \set{s_1, \dots, s_h} \subseteq [k]$, we write $x_S$ for the restriction of $x$ to the entries in $S$. We also write $\hat x_i = x_{[k] \setminus \set{i}}$ for the tuple obtained from~$x$ by omitting~$x_i$. We occasionally write $\MM(a, b, c)$ for the time to multiply an $a \times b$ by a $b \times c$ matrix.

\paragraph{Hyperclique Polynomials}
An \emph{$h$-uniform hypergraph} $G = ([n], E)$ consists of a set $E \subseteq \binom{[n]}{h}$ of \emph{hyperedges} on the vertex set~$[n]$. A \emph{$k$-hyperclique} in~$G$ is a set of $k$ vertices any $h$ of which form a hyperedge. We consider the following algebraic version of this problem: given integer-valued arrays~$A_S$ over $[n]^h$, for all $S \in \binom{[k]}{h}$, the goal is to compute the \emph{$(k, h)$-hyperclique polynomial}
\begin{equation*}
    H_{k, h} = \sum_{x \in [n]^k} \prod_{S \in \binom{[k]}{h}} A_{S; x_S}.
\end{equation*}
If all arrays~$A_S$ are equal to the $0$-$1$-array indicating the hyperedges in~$E$, then $H_{k, h}$ is exactly $k!$ times the number of $k$-hypercliques in~$G$. In particular, $G$ contains a $k$-hyperclique if and only if~$H_{k, h} > 0$. Evaluating $H_{k, h}$ naively takes time $O(n^k)$. We show that $H_{k, h}$ can be evaluated in time $O(n^{(1-\epsilon_h) k})$; this implies \cref{thm:hyperclique}.

\begin{theorem} \label{thm:hyperclique-poly}
For every constant $h \geq 2$, there is a constant $\epsilon_h > 0$ such that, for every constant \makebox{$k > h$}, the polynomial~$H_{k, h}$ can be evaluated in time $O(n^{(1-\epsilon_h) k})$.
\end{theorem}

By the following standard reduction~\cite{NesetrilP85}, it suffices to consider the case $k = h + 1$. Assume that $h+1$ divides~$k$ (otherwise, we first brute-force over $k \bmod (h+1)$ of the variables). We group the $k$ variables into $h+1$ blocks of equal size and view each block as a single variable ranging over~$[n]^{k/(h+1)}$. Each factor $A_{S; x_S}$ depends on only $h$ variables and hence on at most $h$ blocks. We can therefore arrange all factors into $h+1$ arrays such that the $i$-th array does not depend on the $i$-th block. This is an instance of $H_{h+1, h}$ with $n^{k/(h+1)}$ in place of~$n$. Thus, an $O(n^{h+1-\epsilon})$-time algorithm for $H_{h+1, h}$ evaluates $H_{k, h}$ in time $O(n^{(1-\epsilon/(h+1)) k})$. From now on, we therefore focus on the case $k = h + 1$, and our goal is to compute $H_h := H_{h+1, h}$.

\paragraph{Model of Computation}
We work in the word RAM model with words of $\Theta(\log n)$ bits. All integers arising in the algorithms are polynomially bounded (assuming that the input integers are), so arithmetic operations on them take constant time. Having said that, all algorithms only add, subtract, and multiply numbers, so they can also be implemented in algebraic models of computation, e.g., by arithmetic circuits.
\section{Hyperclique in 3-Uniform Hypergraphs}
In this section, we prove \cref{thm:hyperclique} for $3$-uniform hypergraphs. This case already contains all relevant ideas, and it allows us to present them in their simplest form; the next section extends them to all~$h$. Throughout, we stick to basic algebraic language. Our goal is to evaluate the hyperclique polynomial $H_3$, which, for ease of notation, can be stated as follows:
\begin{equation*}
    H_3(A, B, C, D) = \sum_{x, y, z, w \in [n]} A_{xyz} B_{xyw} C_{xzw} D_{yzw}.
\end{equation*}
Evaluating~$H_3$ naively takes time~$O(n^4)$. In this section we show that it can be evaluated in time $O(n^{3.99999})$, thereby proving \cref{thm:hyperclique-poly} for the case $h = 3$. We have made no attempt to optimize constants.

\subsection{Generalized Algebraic Formulation}
What makes the hyperclique polynomial difficult is that each of its variables $x, y, z, w$ appears in three of the four factors. If every variable appeared in only two factors, as in the polynomial
\begin{equation*}
    \sum_{x, y, z, w} A_{xy} B_{yz} C_{zw} D_{wx},
\end{equation*}
we would essentially be counting $4$-cycles in a graph (taking $A = B = C = D$ to be its adjacency matrix). This polynomial is easy to evaluate: it equals $\tr(ABCD)$, which takes two matrix multiplications and thus time $O(n^\omega)$.

For the algorithm we consider a more general polynomial. It lies between the two cases above: we keep the variables $x, y, z, w$, but $A, B, C, D$ become \emph{matrix-valued}. Specifically, the entries in~$A, B, C, D$ become matrices of the following sizes:
\begin{equation*}
    A_{xyz} \in \Int^{a \times c}, \quad B_{xyw} \in \Int^{c \times b}, \quad C_{xzw} \in \Int^{b \times c}, \quad D_{yzw} \in \Int^{c \times a},
\end{equation*}
for all $y, z \in [N]$ and $x, w \in [M]$. Then the \emph{auxiliary hyperclique polynomial} is defined as
\begin{equation*}
    T_3(A, B, C, D) = \sum_{\substack{y, z \in [N]\\x, w \in [M]}} \tr(A_{xyz} B_{xyw} C_{xzw} D_{yzw}).
\end{equation*}
Let us call $(N, M, a, b, c)$ the \emph{size} of this polynomial. This polynomial generalizes the hyperclique polynomial: for $N = M = n$ and $a = b = c = 1$, we have $T_3(A, B, C, D) = H_3(A, B, C, D)$. In contrast, for $N = M = 1$, the polynomial $T_3(A, B, C, D) = \tr(ABCD)$ has exactly the form of the $4$\=/cycle polynomial above and can be evaluated with two matrix multiplications. The algorithm moves from the first case towards the second by gradually decreasing $N, M$ at the cost of increasing~$a, b, c$.

\subsection{Recursive Algorithm}
We describe a simple recursive algorithm to evaluate $T_3$. It repeatedly applies a divide-and-conquer step (\cref{lem:step}) and finally evaluates the resulting polynomials using fast matrix multiplication (\cref{lem:base}).  Throughout, we assume that $N$ and $M$ are powers of some integer constant~$q$ to be chosen later. The recursive step is based on the following simple identity:

\begin{lemma}[Swap Identity] \label{lem:swap}
Let $q \geq 2$, and let $U_{ij} \in \Int^{a \times b}, V_{ij} \in \Int^{b \times a}$ for all \makebox{$i, j \in [q]$}. Then:
\begin{equation*}
    \sum_{i, j \in [q]} \tr(U_{ij} V_{ij}) = \sum_{i, j \in [q]} \tr(U_{ij} V_{ji}) + \sum_{1 \leq i < j \leq q} \tr((U_{ij} - U_{ji})(V_{ij} - V_{ji})).
\end{equation*}
\end{lemma}
\begin{proof}
The terms $i = j$ are the same on both sides, so fix any pair $i < j$. The last trace expands to $\tr(U_{ij} V_{ij}) + \tr(U_{ji} V_{ji}) - \tr(U_{ij} V_{ji}) - \tr(U_{ji} V_{ij})$. The first two terms are the $(i, j)$ and $(j, i)$ terms of the left-hand side. The last two cancel the $(i, j)$ and $(j, i)$ terms of the first sum on the right.
\end{proof}

This identity is not new. For instance, it is exactly the $q \times 2$ Cauchy--Binet formula, extended bilinearly to matrices. The $q = 2$ swap identity has also appeared in~\cite[Appendix~H]{Claude26}, and related identities have shown up for other algorithmic applications, e.g.,~\cite{SolomonikD21}.

\begin{lemma}[Recursive Step] \label{lem:step}
Given input arrays $A, B, C, D$ of size $(N, M, a, b, c)$, we can compute in linear time new arrays $\bar A^g, \bar B^g, \bar C^g, \bar D^g$ for all $g \in [q]$ and~$\bar A^{ij}, \bar B^{ij}, \bar C^{ij}, \bar D^{ij}$ for all $1 \leq i < j \leq q$ such that
\begin{equation*}
    T_3(A, B, C, D) = \sum_{g \in [q]} T_3(\bar A^g, \bar B^g, \bar C^g, \bar D^g) + \sum_{1 \leq i < j \leq q} T_3(\bar A^{ij}, \bar B^{ij}, \bar C^{ij}, \bar D^{ij}).
\end{equation*}
Here, the $q$ polynomials in the first sum have size $(N/q, M, qa, b, c)$ and the $\binom{q}{2}$ polynomials in the second sum have size $(N/q, M, a, b, 2c)$. Alternatively, we can achieve sizes $(N, M/q, a, qb, c)$ and $(N, M/q, a, b, 2c)$ instead.
\end{lemma}
\begin{proof}
We split the variables $y, z \in [N]$ into their last base-$q$ digit and the remaining digits. That is, for all $i, j \in [q]$, we define the arrays $A^{ij}, B^i, C^j, D^{ij}$ by
\begin{align*}
    A^{ij}_{xyz} &= A_{x,\, q(y-1)+i,\, q(z-1)+j}, \\
    B^i_{xyw} &= B_{x,\, q(y-1)+i,\, w}, \\
    C^j_{xzw} &= C_{x,\, q(z-1)+j,\, w}, \\
    D^{ij}_{yzw} &= D_{q(y-1)+i,\, q(z-1)+j,\, w},
\end{align*}
for $y, z \in [N/q]$ and $x, w \in [M]$. Clearly,
\begin{equation*}
    T_3(A, B, C, D) = \sum_{i, j \in [q]} T_3(A^{ij}, B^i, C^j, D^{ij}) = \sum_{i, j \in [q]} \sum_{x, y, z, w} \tr(A^{ij}_{xyz} B^i_{xyw} C^j_{xzw} D^{ij}_{yzw}).
\end{equation*}
These are $q^2$ auxiliary hyperclique polynomials of size $(N/q, M, a, b, c)$, so this alone saves nothing. The savings come from regrouping these terms with \cref{lem:swap}. For now, fix $x, y, z, w$, and for convenience omit them from the subscripts (e.g., \smash{$A^{ij} = A^{ij}_{xyz}$}). Apply \cref{lem:swap} with the matrices~$U_{ij} = A^{ij} B^i$ and $V_{ij} = C^j D^{ij}$. This gives:
\begin{align*}
    \sum_{i, j \in [q]} \tr(A^{ij} B^i C^j D^{ij}) &= \sum_{g \in [q]} \sum_{j \in [q]} \tr(A^{gj} B^g C^g D^{jg}) \\
    &\quad + \sum_{1 \leq i < j \leq q} \tr((A^{ij} B^i - A^{ji} B^j) (C^j D^{ij} - C^i D^{ji})).
\end{align*}
We call the $q$ terms in the first sum the \emph{main terms} (one for each $g \in [q]$) and the $\binom{q}{2}$ terms in the second sum the \emph{correction terms} (one for each $i < j$). We show that each of these terms can be rewritten as a single trace of the form $\tr(\bar A \bar B \bar C \bar D)$.
\begin{itemize}
    \item \emph{Main terms:} After the swap, $B$ and~$C$ only depend on $g$, and the summation over $j$ affects only $A$ and~$D$. We can therefore move~$j$ into the matrix dimension connecting $D$ and~$A$. That is, the $g$-th main term $\sum_{j \in [q]} \tr(A^{gj} B^g C^g D^{jg})$ equals $\tr(\bar A^g \bar B^g \bar C^g \bar D^g)$ for
    \begin{equation*}
        \bar A^g =
        \begin{bmatrix}
            A^{g1} \\
            \vdots \\
            A^{gq} \\
        \end{bmatrix}, \quad
        \bar B^g = B^g, \quad
        \bar C^g = C^g, \quad
        \bar D^g =
        \begin{bmatrix}
            D^{1g} & \cdots & D^{qg}
        \end{bmatrix}\!.
    \end{equation*}
    \item \emph{Correction terms:} Each of the two factors of the $(i, j)$-th correction term is a difference of two products, which we write as a single product of block matrices. That is, the $(i, j)$-th correction term $\tr((A^{ij} B^i - A^{ji} B^j) (C^j D^{ij} - C^i D^{ji}))$ equals $\tr(\bar A^{ij} \bar B^{ij} \bar C^{ij} \bar D^{ij})$ for
    \begin{equation*}
        \bar A^{ij} =
        \begin{bmatrix}
            A^{ij} & -A^{ji} \\
        \end{bmatrix}, \quad
        \bar B^{ij} =
        \begin{bmatrix}
            B^i \\
            B^j \\
        \end{bmatrix}, \quad
        \bar C^{ij} =
        \begin{bmatrix}
            C^j & C^i \\
        \end{bmatrix}, \quad
        \bar D^{ij} =
        \begin{bmatrix}
            D^{ij} \\
            -D^{ji} \\
        \end{bmatrix}\!.
    \end{equation*}
\end{itemize}

Now unfix $x, y, z, w$. The new matrices are chosen in such a way that, e.g., $\bar A^g$ and $\bar A^{ij}$ only depend on the array $A$, and thus only on the indices $x, y, z$. We may therefore view $\bar A^g$ and~$\bar A^{ij}$ as arrays, where the $xyz$-entry is assigned as described before. All in all, we rewrote $T_3(A, B, C, D)$ as:
\begin{align*}
    T_3(A, B, C, D)
    &= \sum_{x, y, z, w} \sum_{i, j \in [q]} \tr(A^{ij}_{xyz} B^i_{xyw} C^j_{xzw} D^{ij}_{yzw}) \\
    &=  \sum_{x, y, z, w} \parens*{\sum_{g \in [q]} \tr(\bar A^g_{xyz} \bar B^g_{xyw} \bar C^g_{xzw} \bar D^g_{yzw}) + \sum_{1 \leq i < j \leq q} \tr(\bar A^{ij}_{xyz} \bar B^{ij}_{xyw} \bar C^{ij}_{xzw} \bar D^{ij}_{yzw})} \\
    &= \vphantom{\Bigg(}\sum_{g \in [q]} T_3(\bar A^g, \bar B^g, \bar C^g, \bar D^g) + \sum_{1 \leq i < j \leq q} T_3(\bar A^{ij}, \bar B^{ij}, \bar C^{ij}, \bar D^{ij}).
\end{align*}
This is the claimed identity. The main terms have size $(N/q, M, qa, b, c)$, since the number of rows of~$\bar A^g$, and equivalently the number of columns of~$\bar D^g$, has increased by a factor of~$q$. Similarly, the correction terms have size $(N/q, M, a, b, 2c)$, as the dimension~$c$ has doubled. Finally, every entry of the new arrays is an entry of $A, B, C, D$, possibly negated, so the new arrays can be constructed in time linear in the input size (recall that $q$ is a constant).

For the alternative sizes, we apply the same construction to the rotated arrays $C, D, A, B$; by the cyclicity of the trace, this exchanges the roles of $y, z$ and $x, w$, and of $a$ and~$b$.
\end{proof}

Note that the recursive step only uses the coefficients $1$ and~$-1$: every entry of the new arrays is an entry of $A, B, C, D$, possibly negated. In particular, the absolute values of the entries never grow throughout the recursion (e.g., if the input arrays are $0$-$1$-arrays, then all arrays in the recursion have entries in $\set{-1, 0, 1}$), so we do not have to worry about integers blowing up in this reduction.

\begin{lemma}[Base Case] \label{lem:base}
The polynomial~$T_3$ of size $(N, N, a, b, c)$ with $a, b, c \leq N$ can be evaluated in time $O(N^4 \cdot (a b c)^{1 - \delta})$ for $\delta = 0.02$.
\end{lemma}
\begin{proof}
We give two separate algorithms to evaluate~$T_3$, which correspond precisely to the two evaluation orders $(AB)(CD)$ and $(BC)(DA)$ from \cref{sec:barriers}. Which of them is faster depends on the dimensions $a$, $b$, and~$c$, and the claimed time bound follows from always executing the faster one.
\begin{itemize}
    \item In the first algorithm, we pair $A$ with $B$ and $C$ with $D$ and evaluate
    \begin{equation*}
        T_3 = \sum_{x, y, z, w \in [N]} \tr((A_{xyz} B_{xyw}) (C_{xzw} D_{yzw})).
    \end{equation*}
    To this end, we compute all the matrices $A_{xyz} B_{xyw}$ and $C_{xzw} D_{yzw}$ for all $x, y, z, w$. Afterwards, we can evaluate~$T_3$ in time $O(N^4 a b)$. To compute the products $A_{xyz} B_{xyw}$, enumerate all $x, y$; omitting~$x, y$ from the subscripts, it remains to compute the products $A_z B_w$ for all $z, w$. We stack the matrices~$A_1, \dots, A_N$ on top of each other, put $B_1, \dots, B_N$ side by side, and compute the product of the resulting large matrices. From this we can read off the desired products $A_z B_w$ for all pairs~$z, w$. The products~$C_{xzw} D_{yzw}$ are computed symmetrically. The running time is dominated by the~$2N^2$ large matrix products. Each has size $Na \times c \times Nb$ and takes time $\MM(Na, c, Nb)$.
    \item In the second algorithm, we instead pair $B$ with $C$ and $D$ with $A$ to evaluate
    \begin{equation*}
        T_3 = \sum_{x, y, z, w \in [N]} \tr((B_{xyw} C_{xzw}) (D_{yzw} A_{xyz})).
    \end{equation*}
    Executing the same strategy amounts to computing~$N^2$ matrix products of size $Nc \times b \times Nc$ plus~$N^2$ matrix products of size $Nc \times a \times Nc$. The final computation of $T_3$ takes time $O(N^4 c^2)$, which is again dominated by the matrix products.
\end{itemize}
The total time is $O\parens{N^2 \min\set{\MM(Na, c, Nb), \MM(Nc, a, Nc) + \MM(Nc, b, Nc)}}$. It suffices to use Strassen's algorithm~\cite{Strassen69}, which multiplies two $d \times d$ matrices in time $O(d^{\log_2 7}) = O(d^{2.81})$. Hence, splitting into square blocks gives $\MM(n, d, m) = O(nm\, d^{0.81})$ whenever $d \leq n, m$ (as is the case here). Since $a^{0.81} + b^{0.81} \leq 2 (ab)^{0.81}$, the total time is $O(N^4 \min\set{ab\, c^{0.81}, (ab)^{0.81} c^2})$, which is $O(N^4 (abc)^{0.98})$ when bounding the minimum by the weighted geometric mean with weights $7/8$ and~$1/8$.
\end{proof}

\begin{proof}[Proof of \cref{thm:hyperclique-poly} for \boldmath$h=3$]
Let $q$ be a sufficiently large integer constant to be chosen later. By padding the arrays with zeros if necessary, we may assume that~$n = q^{2L}$ for some integer~$L$; this increases~$n$ by at most a constant factor. Viewing the input arrays as arrays of $1 \times 1$ matrices, $H_3(A, B, C, D)$ is the auxiliary hyperclique polynomial $T_3(A, B, C, D)$ of size $(n, n, 1, 1, 1)$. We now evaluate this polynomial recursively. 

We apply \cref{lem:step} recursively for $2L$ levels, alternating between the two reductions. Each call of size $(N, M, a, b, c)$ has $q$ \emph{main children} of size~$(N/q, M, qa, b, c)$ and $\binom{q}{2}$ \emph{correction children} of size~$(N/q, M, a, b, 2c)$, or, for the alternative reduction, $q$ main children of size~$(N, M/q, a, qb, c)$ and~$\binom{q}{2}$ correction children of size~$(N, M/q, a, b, 2c)$. By \cref{lem:step}, the polynomial at each internal node is the sum of the polynomials at its children, so~$H_3(A, B, C, D)$ is the sum of the polynomials at the leaves. Each leaf has size $(\sqrt n, \sqrt n, a, b, c)$ with~$a, b \leq q^L = \sqrt n$ and $c \leq 4^L \leq \sqrt n$, so we can evaluate it by \cref{lem:base}.

We analyze the running time. By \cref{lem:step}, each internal node in the recursion tree takes time linear in its input size. Its main children together have the same input size as the node itself, and therefore the total input size can only increase from one recursion level to the next. We can thus charge the running time of the internal nodes to the leaves (each of which takes time at least linear in its input size). Naively, this leads to a negligible overhead of $2L = O(\log n)$, but with a slightly more careful analysis this overhead can be avoided.

It remains to analyze the leaves. Define the \emph{potential} of a node of size $(N, M, a, b, c)$ as $(abc)^{1-\delta}$. The root has potential~$1$, and the children of a node with potential~$\Phi$ have total potential
\begin{equation*}
    q \cdot q^{1-\delta} \Phi + \binom{q}{2} \cdot 2^{1-\delta} \Phi;
\end{equation*}
this becomes $\rho q^2 \Phi$ for
\begin{equation*}
    \rho = q^{-\delta} + (1 - 1/q)\, 2^{-\delta} \leq 2^{-\delta} (1 + (2/q)^{\delta}).
\end{equation*}
Hence, the leaves have total potential $(\rho q^2)^{2L} = n^2 \rho^{2L}$. By \cref{lem:base}, a leaf of potential~$\Phi$ takes time $O((\sqrt n)^4 \Phi) = O(n^2 \Phi)$, so all leaves together take time $O(n^4 \rho^{2L})$. Recalling that $L = \log_q(n)/2$, this becomes $O(n^{4 - \epsilon})$ for
\begin{equation*}
    \epsilon = \log_q(1/\rho) \geq \frac{\delta - \log_2(1 + (2/q)^{\delta})}{\log_2 q}.
\end{equation*}
This shows the trade-off in choosing~$q$: it must be large enough to make $(2/q)^{\delta}$ small, but $\epsilon$ decays with~$\log_2 q$. We choose $q = 2^{400}$. Then, as $\delta = 0.02$, we have $(2/q)^{\delta} < 2^{-7}$ and $\log_2(1 + 2^{-7}) < 0.012$, and thus $\epsilon > 0.008/400 = 2 \cdot 10^{-5}$. The total running time is $O(n^{3.99998} \log n) = O(n^{3.99999})$.
\end{proof}

As mentioned before, we have made no attempt to optimize the constants. A simple optimization is to rely on the state-of-the-art bounds for square and rectangular matrix multiplication~\cite{OpenAI26a,Uznanski26}. Analyzing the recursion more carefully, one can then obtain time~$O(n^{3.9583})$, but it is likely that much more dramatic improvements are possible.

\section{Hyperclique in \texorpdfstring{\boldmath$h$}{h}-Uniform Hypergraphs}
In this section we prove \cref{thm:hyperclique-poly} in full generality. To this end, we generalize the algorithm from the previous section from computing $H_3$ to computing $H_h$ for all $h \geq 2$. The proof is by induction on~$h$. In the base case, for $h = 2$, the problem degenerates to matrix multiplication, so that, using Strassen's algorithm, $H_2$ can be evaluated in time $O(n^{3-\epsilon_2})$ for $\epsilon_2 := 3 - \log_2 7 > 0$.

For the inductive step, let $h \geq 3$, and assume that $H_{h-1}$ can be evaluated in time $O(n^{h-\epsilon_{h-1}})$ for some constant $0 < \epsilon_{h-1} \leq \epsilon_2$ (decreasing~$\epsilon_{h-1}$ only weakens this assumption); the goal is to evaluate~$H_h$ in time $O(n^{h+1-\epsilon_h})$ for some constant $\epsilon_h > 0$. The rest of this section focuses on this inductive step. As in the previous section, it is based on a recursive algorithm that evaluates a more general polynomial~$T_h$. Since this closely follows the previous section, we focus on the technical details and only comment on aspects that are new. Throughout, $h$ is a constant, and the $O$-notation hides all factors depending on~$h$. For convenience, we express $H_h$ as
\begin{equation*}
    H_h(A, B, C, D, E_5, \dots, E_{h+1}) = \sum_{x \in [n]^{h+1}} A_{\hat x_4} B_{\hat x_3} C_{\hat x_2} D_{\hat x_1} \prod_{\ell=5}^{h+1} E_{\ell; \hat x_\ell}.
\end{equation*}
In this view the arrays $A, B, C, D$ play the same role as before; the arrays $E_5, \dots, E_{h+1}$ need extra care.

To define~$T_h$, again let $A, B, C, D$ be matrix-valued arrays, where each matrix has size
\begin{equation*}
    A_{\hat x_4} \in \Int^{a \times c}, \quad B_{\hat x_3} \in \Int^{c \times b}, \quad C_{\hat x_2} \in \Int^{b \times c}, \quad D_{\hat x_1} \in \Int^{c \times a},
\end{equation*}
as in the previous section. In addition, the arrays $E_5, \dots, E_{h+1}$ have three-dimensional entries:
\begin{equation*}
    E_{5; \hat x_5}, \dots, E_{h+1; \hat x_{h+1}} \in \Int^{a \times b \times c}.
\end{equation*}
Abbreviating $E = (E_5, \dots, E_{h+1})$, the \emph{auxiliary hyperclique polynomial} is
\begin{equation*}
    T_h(A, B, C, D, E) = \sum_x \sum_{\substack{\alpha \in [a], \beta \in [b]\\\gamma \in [c]}} (A_{\hat x_4} B_{\hat x_3})[\alpha, \beta] \cdot C_{\hat x_2}[\beta, \gamma] \cdot D_{\hat x_1}[\gamma, \alpha] \cdot \prod_{\ell=5}^{h+1} E_{\ell; \hat x_\ell}[\alpha, \beta, \gamma].
\end{equation*}
We call $(N, M, a, b, c)$ its \emph{size}, where, as before, $x_2, x_3$ range over~$[N]$ and $x_1, x_4$ range over~$[M]$; the variables $x_5, \dots, x_{h+1}$ always range over~$[n]$. For $N = M = n$ and $a = b = c = 1$, we have $T_h = H_h$. For $h = 3$, there are no arrays~$E_\ell$, and $T_h = \sum_x \tr(A_{\hat x_4} B_{\hat x_3} C_{\hat x_2} D_{\hat x_1})$ is exactly the polynomial~$T_3$ from the previous section. 

The entries of the arrays~$E_\ell$ are three-dimensional for the following reason. Each~$E_\ell$ depends on all variables $x_1, \dots, x_4$ that the recursive step splits. Therefore, it has to follow both the main terms, which move digits of these variables into the dimensions~$a$ and~$b$, and the correction terms, which double the dimension~$c$. Its three indices $\alpha$, $\beta$, and~$\gamma$ correspond exactly to these three dimensions.

\subsection{Recursive Algorithm}
We now describe the recursive algorithm to evaluate~$T_h$. As before, we assume that $N$ and $M$ are powers of some integer constant~$q$ to be chosen later.

\begin{lemma}[Recursive Step] \label{lem:step-general}
Given arrays $A, B, C, D, E$ of size $(N, M, a, b, c)$, we can compute in linear time new arrays $\bar A^g, \bar B^g, \bar C^g, \bar D^g, \bar E^g$ for all $g \in [q]$ and $\bar A^{ij}, \bar B^{ij}, \bar C^{ij}, \bar D^{ij}, \bar E^{ij}$ for all $1 \leq i < j \leq q$ such that
\begin{equation*}
    T_h(A, B, C, D, E) = \sum_{g \in [q]} T_h(\bar A^g, \bar B^g, \bar C^g, \bar D^g, \bar E^g) + \sum_{1 \leq i < j \leq q} T_h(\bar A^{ij}, \bar B^{ij}, \bar C^{ij}, \bar D^{ij}, \bar E^{ij}).
\end{equation*}
Here, the polynomials in the first sum have size $(N/q, M, qa, b, c)$ and those in the second sum have size $(N/q, M, a, b, 2c)$. Alternatively, we can achieve sizes $(N, M/q, a, qb, c)$ and $(N, M/q, a, b, 2c)$ instead.
\end{lemma}
\begin{proof}
We start with the first reduction. As in the proof of \cref{lem:step}, we split $x_2, x_3 \in [N]$ into their last base-$q$ digit and the remaining digits. This yields arrays $A^{ij}, B^i, C^j, D^{ij}$ as before and, since every~$E_\ell$ depends on both $x_2$ and~$x_3$, arrays $E^{ij}_\ell$ such that
\begin{equation*}
    T_h(A, B, C, D, E) = \sum_{i, j \in [q]} T_h(A^{ij}, B^i, C^j, D^{ij}, E^{ij}).
\end{equation*}
In the previous section, we applied \cref{lem:swap} to the matrices $A^{ij} B^i$ and $C^j D^{ij}$. Here, we also have to take the arrays $E^{ij}_\ell$ into account, and we simply attach them to the second factor. For now, fix~$x$, and for convenience omit it from the subscripts (e.g., \smash{$A^{ij} = A^{ij}_{\hat x_4}$}). We again let $U_{ij} = A^{ij} B^i$, and we let $V_{ij} \in \Int^{b \times a}$ be the matrix with entries
\begin{equation*}
    V_{ij}[\beta, \alpha] = \sum_{\gamma \in [c]} C^j[\beta, \gamma] \cdot D^{ij}[\gamma, \alpha] \cdot \prod_{\ell=5}^{h+1} E^{ij}_\ell[\alpha, \beta, \gamma].
\end{equation*}
Without the arrays~$E_\ell$, this would be exactly the product~$C^j D^{ij}$. Since $T_h(A^{ij}, B^i, C^j, D^{ij}, E^{ij})$ is the sum of $\tr(U_{ij} V_{ij})$ over all~$x$, we can apply \cref{lem:swap} as before. For $A, B, C, D$, we use the same block matrices as in the proof of \cref{lem:step}. In the $g$-th main term, the remaining digit~$j$ appears in~$A$, $D$, and all~$E_\ell$, and we move it into the dimension~$a$ by setting $\bar E^g_\ell[(j, \alpha), \beta, \gamma] = E^{jg}_\ell[\alpha, \beta, \gamma]$, where we identify $[q] \times [a]$ with $[qa]$ as for the rows of~$\bar A^g$. In the $(i, j)$-th correction term, we set $\bar E^{ij}_\ell[\alpha, \beta, (1, \gamma)] = E^{ij}_\ell[\alpha, \beta, \gamma]$ and $\bar E^{ij}_\ell[\alpha, \beta, (2, \gamma)] = E^{ji}_\ell[\alpha, \beta, \gamma]$, where we identify $[2] \times [c]$ with $[2c]$ as for the columns of~$\bar C^{ij}$. Then:
\begin{align*}
    \sum_{j \in [q]} \tr(U_{gj} V_{jg})
    &= \sum_{\alpha', \beta, \gamma} (\bar A^g \bar B^g)[\alpha', \beta] \cdot \bar C^g[\beta, \gamma] \cdot \bar D^g[\gamma, \alpha'] \cdot \prod_{\ell=5}^{h+1} \bar E^g_\ell[\alpha', \beta, \gamma], \\
    \tr((U_{ij} - U_{ji}) (V_{ij} - V_{ji}))
    &= \sum_{\alpha, \beta, \gamma'} (\bar A^{ij} \bar B^{ij})[\alpha, \beta] \cdot \bar C^{ij}[\beta, \gamma'] \cdot \bar D^{ij}[\gamma', \alpha] \cdot \prod_{\ell=5}^{h+1} \bar E^{ij}_\ell[\alpha, \beta, \gamma'],
\end{align*}
where $\alpha' = (j, \alpha)$ ranges over $[qa]$ and $\gamma' = (t, \gamma)$ ranges over $[2c]$. Now unfix~$x$. As in the previous section, each new matrix only depends on the indices of the array it is formed from; for instance, $\bar A^g$ and $\bar A^{ij}$ only depend on~$\hat x_4$, and $\bar E^g_\ell$ and $\bar E^{ij}_\ell$ only depend on~$\hat x_\ell$. We may therefore view the new matrices as arrays indexed by $x$, and summing the above identities over all~$x$ yields exactly~$T_h(\bar A^g, \bar B^g, \bar C^g, \bar D^g, \bar E^g)$ and $T_h(\bar A^{ij}, \bar B^{ij}, \bar C^{ij}, \bar D^{ij}, \bar E^{ij})$. Every entry of the new arrays is an entry of the old arrays, possibly negated, so the construction takes linear time.

The second reduction is analogous: we split $x_1$ and $x_4$ instead, which are shared by $B$, $C$, and all~$E_\ell$. In the main terms, the remaining digit then appears in $B$, $C$, and all~$E_\ell$, and we move it into the dimension~$b$, that is, into the index~$\beta$. The correction terms are handled as before.
\end{proof}

The base case uses the induction hypothesis in the following form.

\begin{lemma}[Rectangular Hyperclique] \label{lem:rectangular}
Assume that $H_{h-1}$ can be evaluated in time $O(n^{h-\epsilon_{h-1}})$ for some~\makebox{$\epsilon_{h-1} > 0$}. Then every instance of~$H_{h-1}$ in which the $h$ variables range over $[n_1], \dots, [n_h]$, respectively, can be evaluated in time
\begin{equation*}
    O\parens*{\frac{n_1 \cdot \ldots \cdot n_h}{\min\set{n_1, \dots, n_h}^{\epsilon_{h-1}}}}.
\end{equation*}
\end{lemma}
\begin{proof}
Let $m = \min\set{n_1, \dots, n_h}$. Split each range into blocks of size~$m$, padding with zeros if necessary. This splits the instance into $O((n_1/m) \ldots (n_h / m))$ instances of $H_{h-1}$ in which all variables range over~$[m]$. Each can be constructed in linear time and evaluated in time $O(m^{h-\epsilon_{h-1}})$.
\end{proof}

\begin{lemma}[Base Case] \label{lem:base-general}
Assume that $H_{h-1}$ can be evaluated in time $O(n^{h-\epsilon_{h-1}})$ for some~\makebox{$\epsilon_{h-1} > 0$}. Then $T_h$ of size $(N, N, a, b, c)$ with $a, b, c \leq N \leq n$ can be evaluated in time $O(N^4 n^{h-3} (abc)^{1-\delta})$ for~$\delta = \epsilon_{h-1}^2/8$.
\end{lemma}
\begin{proof}
We give three algorithms and run the fastest one. The main difference to the previous section is that, after fixing some of the variables and indices, the remaining sum is again a hyperclique polynomial (of uniformity $h-1$), which we evaluate using \cref{lem:rectangular} instead of matrix multiplication.
\begin{itemize}
    \item The first algorithm pairs $A$ with~$B$, and then evaluates the remaining sum using \cref{lem:rectangular}. In detail, we compute the products $W_x = A_{\hat x_4} B_{\hat x_3}$ for all~$x$. This amounts to $N^2 n^{h-3}$ matrix products of size $Na \times c \times Nb$, one for each choice of the variables $x_1, x_2, x_5, \dots, x_{h+1}$ shared by $A$ and~$B$. Using Strassen's algorithm, this takes time \smash{$O(N^4 n^{h-3} ab c^{1 - \epsilon_2}) \leq O(N^4 n^{h-3} ab c^{1-\epsilon_{h-1}})$}, as we are assuming $\epsilon_{h-1} \leq \epsilon_2$. Afterwards, we fix $x_3, x_4 \in [N]$, $\alpha \in [a]$, and $\beta \in [b]$. The remaining sum
    \begin{equation*}
        \sum_{\gamma, x_1, x_2, x_5, \dots, x_{h+1}} W_x[\alpha, \beta] \cdot C_{\hat x_2}[\beta, \gamma] \cdot D_{\hat x_1}[\gamma, \alpha] \cdot \prod_{\ell=5}^{h+1} E_{\ell; \hat x_\ell}[\alpha, \beta, \gamma]
    \end{equation*}
    is an instance of $H_{h-1}$ with the $h$ variables $\gamma$, $x_1$, $x_2$, and $x_5, \dots, x_{h+1}$, since its $h$ factors omit $\gamma$, $x_2$, $x_1$, and $x_5, \dots, x_{h+1}$, respectively. By \cref{lem:rectangular}, it can be evaluated in time~$O(c^{1-\epsilon_{h-1}} N^2 n^{h-3})$. Summing over all $N^2 ab$ choices, the first algorithm takes time $O(N^4 n^{h-3} ab c^{1-\epsilon_{h-1}})$ in total.
    \item The second algorithm instead pairs $B$ with~$C$, so that the index~$\alpha$ becomes one of the variables of the remaining hyperclique polynomial. Expand $(A_{\hat x_4} B_{\hat x_3})[\alpha, \beta] = \sum_{\kappa \in [c]} A_{\hat x_4}[\alpha, \kappa] \cdot B_{\hat x_3}[\kappa, \beta]$ in the definition of~$T_h$, and fix $x_2, x_3 \in [N]$, $\beta \in [b]$, and $\gamma, \kappa \in [c]$. It remains to compute
    \begin{equation*}
        \sum_{\alpha, x_1, x_4, x_5, \dots, x_{h+1}} A_{\hat x_4}[\alpha, \kappa] \cdot (B_{\hat x_3}[\kappa, \beta] \cdot C_{\hat x_2}[\beta, \gamma]) \cdot D_{\hat x_1}[\gamma, \alpha] \cdot \prod_{\ell=5}^{h+1} E_{\ell; \hat x_\ell}[\alpha, \beta, \gamma].
    \end{equation*}
    This remaining sum is an instance of $H_{h-1}$ with the $h$ variables $\alpha$, $x_1$, $x_4$, and $x_5, \dots, x_{h+1}$, since its $h$ factors omit $x_4$, $\alpha$, $x_1$, and $x_5, \dots, x_{h+1}$, respectively, viewing $B_{\hat x_3}[\kappa, \beta] \cdot C_{\hat x_2}[\beta, \gamma]$ as one factor. By \cref{lem:rectangular}, it can be evaluated in time $O(a^{1-\epsilon_{h-1}} N^2 n^{h-3})$. Summing over all $N^2 b c^2$ choices of $x_2, x_3, \beta, \gamma, \kappa$, the second algorithm takes time $O(N^4 n^{h-3} b c^2 a^{1-\epsilon_{h-1}})$.
    \item The third algorithm is symmetric to the second one and pairs $D$ with~$A$, so that the index~$\beta$ becomes one of the variables of the remaining hyperclique polynomial. We use the same expansion of $(A_{\hat x_4} B_{\hat x_3})[\alpha, \beta]$, but now fix $x_1, x_4 \in [N]$, $\alpha \in [a]$, and $\gamma, \kappa \in [c]$. It remains to compute
    \begin{equation*}
        \sum_{\beta, x_2, x_3, x_5, \dots, x_{h+1}} (D_{\hat x_1}[\gamma, \alpha] \cdot A_{\hat x_4}[\alpha, \kappa]) \cdot B_{\hat x_3}[\kappa, \beta] \cdot C_{\hat x_2}[\beta, \gamma] \cdot \prod_{\ell=5}^{h+1} E_{\ell; \hat x_\ell}[\alpha, \beta, \gamma].
    \end{equation*}
    This is an instance of $H_{h-1}$ with the $h$ variables $\beta$, $x_2$, $x_3$, and $x_5, \dots, x_{h+1}$, since its $h$ factors omit $\beta$, $x_3$, $x_2$, and $x_5, \dots, x_{h+1}$, respectively, viewing $D_{\hat x_1}[\gamma, \alpha] \cdot A_{\hat x_4}[\alpha, \kappa]$ as one factor. By \cref{lem:rectangular}, it can be evaluated in time $O(b^{1-\epsilon_{h-1}} N^2 n^{h-3})$. Summing over all $N^2 a c^2$ choices of $x_1, x_4, \alpha, \gamma, \kappa$, the third algorithm takes time $O(N^4 n^{h-3} a c^2 b^{1-\epsilon_{h-1}})$.
\end{itemize}
The fastest of the three algorithms takes time $O(N^4 n^{h-3} \min\set{a b c^{1-\epsilon_{h-1}}, a^{1-\epsilon_{h-1}} b c^2, a b^{1-\epsilon_{h-1}} c^2})$. Bounding the minimum by a weighted geometric mean with weights $1 - \epsilon_{h-1}/4$, $\epsilon_{h-1}/8$, and $\epsilon_{h-1}/8$, we obtain
\begin{equation*}
    \min\set{a b c^{1-\epsilon_{h-1}}, a^{1-\epsilon_{h-1}} b c^2, a b^{1-\epsilon_{h-1}} c^2} \leq (ab)^{1-\epsilon_{h-1}^2/8} c^{1-3\epsilon_{h-1}/4+\epsilon_{h-1}^2/4} \leq (abc)^{1-\epsilon_{h-1}^2/8},
\end{equation*}
and hence, the fastest algorithm runs in time $O(N^4 n^{h-3} (abc)^{1-\delta})$ for $\delta = \epsilon_{h-1}^2 / 8$.
\end{proof}

\begin{proof}[Proof of \cref{thm:hyperclique-poly}]
The proof is by induction on $h$. Recall that in the base case $h = 2$, evaluating $H_2$ degenerates to computing one matrix product. By, e.g., Strassen's algorithm, this proves the claim for $\epsilon_2 = 3 - \log_2 7 > 0$.

Next, focus on the inductive step for $h \geq 3$, i.e., assume that $H_{h-1}$ can be evaluated in time $O(n^{h-\epsilon_{h-1}})$ by the induction hypothesis; we give an algorithm to evaluate $H_h$. Let $\delta = \epsilon_{h-1}^2/8$ and let $q$ be an integer to be determined later. We pad~$n$ to \smash{$q^{2L}$} for some integer~$L$. Then we view $H_h$ as an auxiliary hyperclique polynomial $T_h$ of size $(n, n, 1, 1, 1)$, and design a recursive algorithm to evaluate $T_h$. Apply \cref{lem:step-general} recursively for $2L$ levels, alternating between the two reductions. Thereby, in the recursion tree, each internal node of size $(N, M, a, b, c)$ has $q$ main children of size $(N/q, M, qa, b, c)$ and $\binom{q}{2}$ correction children of size $(N/q, M, a, b, 2c)$, or alternatively $q$ main children of size $(N, M/q, a, qb, c)$ and $\binom{q}{2}$ correction children of size $(N, M/q, a, b, 2c)$. Each leaf has size $(\sqrt n, \sqrt n, a, b, c)$ with \makebox{$a, b, c \leq \sqrt n$}, so we can evaluate it by \cref{lem:base-general}.

Since the main children of each node together have the same input size as the node, the internal nodes worsen the running time by at most a factor $O(\log n)$. To analyze the total time of the leaves, associate the \emph{potential} $(abc)^{1-\delta}$ to any node of size $(N, M, a, b, c)$. A leaf of potential~$\Phi$ takes time~$O(n^{h-1} \Phi)$, and by the same calculation as before, all leaves together take time $O(n^{h+1-\epsilon})$ for
\begin{equation*}
    \epsilon \geq \frac{\delta - \log_2(1 + (2/q)^{\delta})}{\log_2 q}.
\end{equation*}
For $\epsilon$ to be positive, $q$ must be large enough to make $(2/q)^{\delta}$ small, but then $\epsilon$ decays with~$\log_2 q$. We therefore set $q = 2^{\ceil{(2/\delta) \log_2(1/\delta)}}$. Since $\delta \leq 1/8$, we have $(2/q)^{\delta} = 2^{-\delta(\log_2 q - 1)} \leq 2^{\delta} \delta^2 \leq \delta/4$, and thus $\log_2(1 + (2/q)^{\delta}) \leq \delta/2$. Together with $\log_2 q \leq (3/\delta) \log_2(1/\delta)$, this gives \makebox{$\epsilon \geq \delta^2/(6 \log_2(1/\delta))$}. Hence, $H_h$ can be evaluated in time $O(n^{h+1-\epsilon} \log n) = O(n^{h+1-\epsilon_h})$ for $\epsilon_h = \delta^2/(12 \log_2(1/\delta)) > 0$. As a final remark, note that the recursive step only negates entries, and all numbers in the algorithm remain polynomially bounded for polynomially bounded integer inputs.
\end{proof}

The recursion $\epsilon_h = \Omega(\epsilon_{h-1}^4 / \log(1/\epsilon_{h-1}))$ shows that $\epsilon_h$ decays doubly exponentially in~$h$. We have made no attempt to optimize this dependence.

\section*{Acknowledgments}
I would like to thank Marvin Künnemann for pointing me to some relevant references.

\bibliographystyle{alphaurl}
\bibliography{refs}

\end{document}